# Prototyping coronagraphs for exoplanet characterization with SPHERE

Anthony Boccaletti[a], Lyu Abe[b], Jacques Baudrand[a], Jean-Baptiste Daban[b], Richard Douet[b], Geraldine Guerri[b], Sylvie Robbe-Dubois[b], Philippe Bendjoya[b], Kjetil Dohlen[c], and Dimitri Mawet[d*]

[a] LESIA, Observatoire de Paris, 5 pl. J. Janssen, F-92195 Meudon, France;
[b] FIZEAU, Université de Nice-Sophia Antipolis, Parc Valrose, F-06108 Nice, France
[c] LAM, Observatoire de Marseille-Provence, 38 rue F. Joliot-Curie, F-13388 Marseille, France
[d] JPL, 4800 Oak Grove Drive, Pasadena, California 91109, US

## ABSTRACT

The detection and characterization of extrasolar planets with SPHERE (Spectro Polarimetric High contrast Exoplanet REsearch) is challenging and in particular relies on the ability of a coronagraph to attenuate the diffracted starlight. SPHERE includes 3 instruments, 2 of which can be operated simultaneously in the near IR from 0.95 to 1.8 microns. This requirements is extremely critical for coronagraphy. This paper briefly introduces the concepts of 2 coronagraphs, the Half-Wave Plate Four Quadrant Phase Masks and the Apodized Pupil Lyot Coronagraph, prototyped within the SPHERE consortium by LESIA (Observatory of Paris) and FIZEAU (University of Nice) respectively. Then, we present the measurements of contrast and sensitivity analysis. The comparison with technical specifications allows to validate the technology for manufacturing these coronagraphs.

**Keywords:** Coronagraphy, High Contrast Imaging, Exoplanets,

## 1. INTRODUCTION

The understanding of the planetary formation stage has triggered a large interest for the detection and the characterization of exoplanets since a few decades. However, for now, indirect techniques have been the most productive to found planetary mass objects. Radial velocity survey of bright stars, microlensing in the galactic plane or transit monitoring with modest telescopes have provided a significant catalogue of more than 250 planetary mass objects. The aforesaid techniques have the ability to derive mass, radius, and sometimes temperature estimates but the spectroscopic characterization is still pending direct imaging techniques except for a very few examples like HD189733 (Tinetti et al. 2007).

By the end of 2010, SPHERE (Beuzit et al. 2006) at the VLT will have the capability to contribute to this field of research as GPI (Macintosh et al. 2006) at Gemini. SPHERE is the Spectro-Polarimetric High-contrast Exoplanet REsearch designed to obtain spectra of extrasolar giant gaseous planets in the near IR.

SPHERE makes use of differential imaging to detect the signal of gaseous planets in the near IR. This concept is implemented as a dual beam camera (IRDIS) and an integral field spectrograph (IFS) for the near IR part and as a dual polarization camera in the visible (ZIMPOL).

For 80% of the time SHERE will operate in a survey mode in which IRDIS and IFS observe simultaneously thus covering a large bandwidth for the Y to H band (0.95-1.80μm). This particular observing mode is favorable for the detection of exoplanets as differential imaging will benefit for large bandwidths (Sparks & Ford 2002) and false alarms are reduce by the simultaneity. However, it puts strong constraints on chromaticity issues, especially for the starlight suppression device.

With the emergence of projects geared towards the direct detection of exoplanets the development of high contrast imaging techniques has became predominant and in particular coronagraphy. The purpose of coronagraphy is to attenuate the on-axis object while leaving unchanged the light of off-axis object. Since the invention by B. Lyot in the 30's to observe the Solar corona (Lyot 1939), the concept has a lot evolved (phase masks, improved Lyot designs, or interferometric concepts). A comprehensive description of the many concepts is given in Guyon et al. 2006. For SPHERE, two concepts were identified to address the problem of chromaticity. The first one is an achromatic version of the 4 Quadrant Phase Mask (4QPM, Rouan et al. 2000) developed at LESIA and the second one is the Apodized Pupil Lyot Coronagraph developed at FIZEAU (APLC, Aime et al. 2002).

---

[*] Further author information: (Send correspondence to A.B.)
A.B.: E-mail: anthony.boccaletti@obspm.fr, Telephone: +33 (0)1 45 07 77 21

This paper describes the two prototypes built to demonstrate the feasibility of achromatic coronagraphs suited for SPHERE and presents a few results of an intensive testing phase in 2007.

## 2. CORONAGRAPH PROTOTYPES

### 2.1 The HW-4QPM

The 4QPM was the subject of intensive studies in terms of numerical simulations as well as lab experiments. The LESIA is pursuing an active development of the 4QPM in different wavelength regimes for different projects. First laboratory experiment in the visible provided an attenuation of the central peak by a factor $10^5$ and contrast level of $10^{-6}$ at $3\lambda/D$ (Riaud et al. 2003). Near IR 4QPMs are installed in NACO at the VLT (Boccaletti et al. 2004) and are able to deliver useful scientific data (Gratadour et al. 2005, Riaud et al. 2006). Several mid-IR 4QPMs will be included in MIRI the mid IR instrument of JWST (Baudoz et al. 2006).

The greatest disadvantage of the 4QPMs is the chromaticity. For NACO and MIRI, the chromatism was not an issue either because the spectral bandwidth is narrow or because the level of aberrations is large. In SPHERE the requirements are much stringent especially in terms of spectral range and a coronagraph is needed to feed at the same time the two near IR instruments IRDIS and IFS.

The Half-Waveplates 4QPM is an achromatic version of the original 4QPM design. It relies on the properties of some materials, the so-called birefringence. Achromatic waveplates are commercially manufactured using a combination of two birefringent materials in order to make the optical path difference linear with lambda across a given. To mimic the 4QPM design, four identical half-waveplates are assembled side-by-side, two of them being rotated by 90° around their normal as shown in Mawet et al. 2006. This component, the HW-4QPM, is behaving as two achromatic 4QPMs, one for each polarization state. In theory, residual phase can be relatively small in the spectral range considered ($10^{-2}$ - $10^{-1}$ radian). The difficulty relies in the manufacturing and the assembling. The present prototype makes use of Quartz and $MgF_2$ as it was already the case for a former prototype adapted to the visible spectrum and tested at the Observatoire de Meudon (Mawet et al. 2006). The goal here, is to demonstrate the feasibility for the near IR in the SPHERE bandwidth and to assess the sensibility of this component to various parameters.

A schematic of the prototype is shown in Figure 1 and an image of the actual assembly in Figure 2.

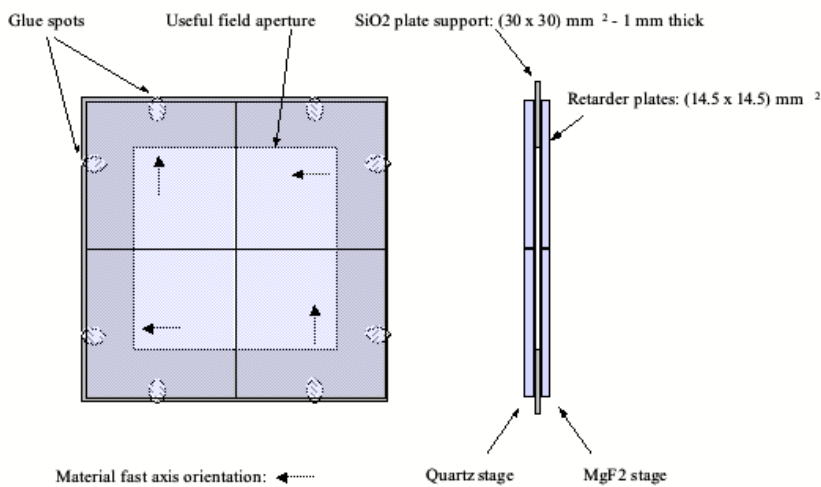

Figure 1: schematic of the HW-4QPM mask seen from the top (left) and from the side (right). The intermediate plate is a polished substrate with a hole delimiting the FOV of about 20x20mm. Arrows indicate the direction of a polarization state.

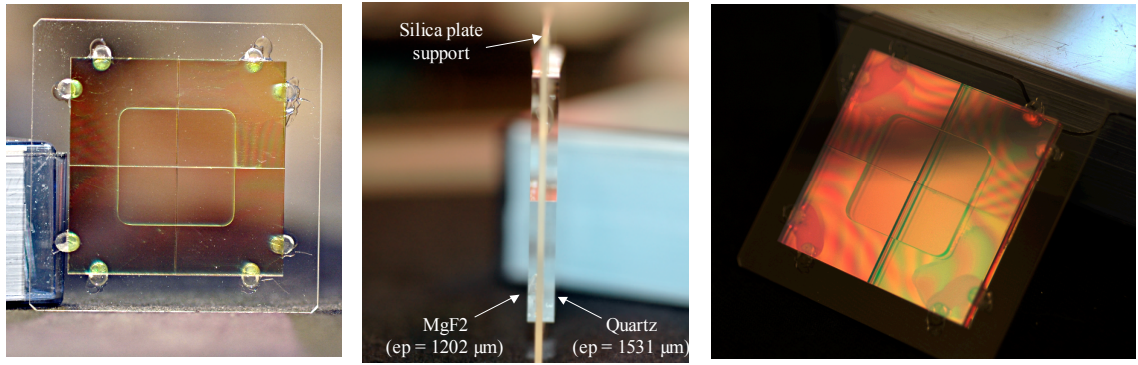

Figure 2: Pictures of the HW-4QPM prototype. Front and rear view (left and centre) and tilted view (right) show fringe pattern allowing control of the parallelism between the waveplates

The specifications of the prototype are given in Table 1. The alignment of the two stages is obtained with an accuracy of about 10μm. The co-planeity of the waveplates is reached with an accuracy of 10 arcmin. The theoretical coronagraphic extinction of this component based on chromaticity leakage is larger than $10^3$ accross the whole bandwidth.

Table 1: Technical specification for the HW-4QPM prototype.

| Manufacturing specifications | Value |
|---|---|
| Spectral Range | 950 - 1800 nm |
| Side of each waveplates | 14.5 ± 0.1 mm |
| Angle of each waveplates | 90° ± 15" |
| Materials | Quartz / MgF2 |
| Quartz thickness | 1668.0 ± 0.2 mm |
| MgF$_2$ thickness | 1305.5 ± 0.5mm |
| Cutting accuracy | 5 mm |
| Optical surface quality | λ/10 PTV @ 633 nm |
| Parallelism | < 5 arcsec |
| Crystal axis w.r.t. waveplate edges | < 5 arcmin |
| Anti-Reflection Coating for (950-1800 nm) | R < 0.5% both sides |

## 2.2 The APLC

One of the limitations that occur with a Lyot coronagraph is the light distribution in the relayed pupil. This distribution is dictated by the convolution of the telescope aperture with the Fourier transform of the mask. To reduce the contamination of the starlight inside the geometric aperture, the Lyot mask have to be quite large and the Lyot stop smaller, a condition which is generally not favorable when looking at faint objects like exoplanets near bright stars. A solution to improve the Lyot design is to reduce the oscillation of the mask Fourier Transform. It is achieved with the Apodization or literally "attenuation of the PSF feet". An Apodizer is a component with a non-uniform transmission located in the entrance pupil of the system. Although the concept of apodization is quite old, its combination with Lyot coronagraph was proposed in Baudoz et al. 2000 and also Guyon & Roddier 2000. The exact mathematical solutions of the apodizer was found by Aime et al. 2002 for rectangular apertures and later on for any aperture (Soummer 2005) and belongs to the family of Prolate functions. The particularity of Prolate functions is that their Fourier Transform is a truncated Prolate function and conversely. A coronagraph makes the interference (or subtraction) of two waves in the relayed pupil: the aperture and Fourier Transform of the mask. With an apodizer, the aperture is made closer to the Fourier transform of the mask and the subtraction is therefore more efficient than with a conventional Lyot coronagraph. Trade-off with respect to the mask size indicates an optimal diameter of 4λ/D which is defined for a wavelength of 1.65μm. The development of the APLC prototype for SPHERE is led by FIZEAU. The prototype apodizer has been purchased in phase B to Reynard Corp. in order to confirm the feasibility of this component and to measure its coronagraphic performance on a test bench. The technology of this apodizer is a thin film of Inconel® 600 on a glass substrate. Inconel® 600 has two advantages: a pretty low reflectivity coefficient and a flat spectral transmission in visible and NIR wavelengths. These features will be analyzed in the next paragraphs. Figure 3 shows the apodizer profile with its tolerance limits and specifications are given in Table 2.

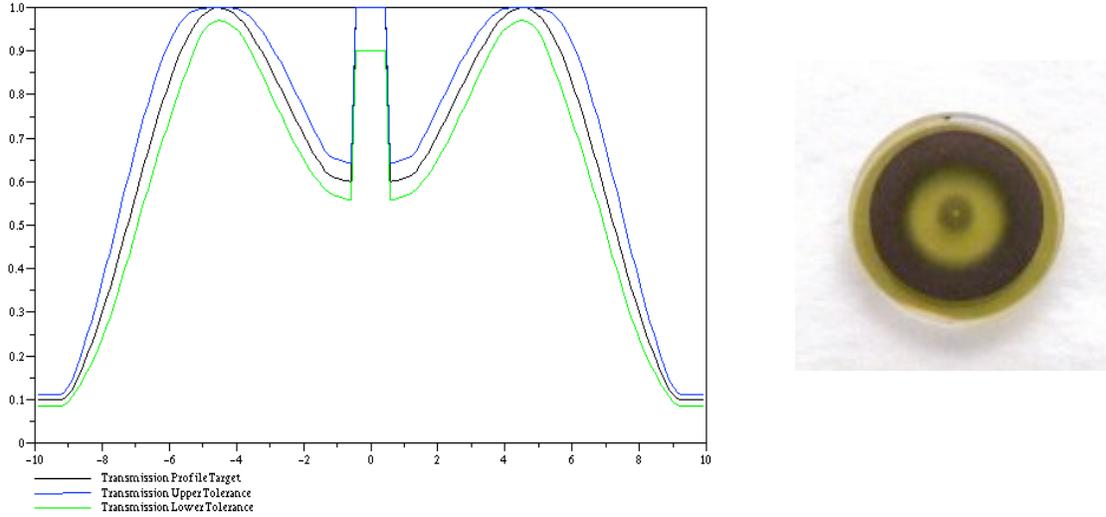

Figure 3: transmission profile of the apodizer (left) and picture of the apodizer prototype designed for an 18mm pupil (right).

Table 2: Technical specification for the APLC prototype.

| Manufacturing specification | Value |
|---|---|
| Spectral Range | 950 nm – 1800 nm |
| Size | |
|     Outer diameter of the Inconel coated area: OD | 19.8mm±0.1mm |
|     Inner diameter of the Inconel coated area: ID | 1.0mm ±0.05mm |
| Transmission profile | |
|     Profile is circular symmetric | |
|     Profile's transmission: | see Fig. 3 |
| Outside diameter of the substrate | 30±0.1mm |
| Thickness of the substrate | 4±0.1mm |
| Optical surface quality | $\lambda/10$ PTV @633 nm |
| Parallelism | <5 arcsec |
| Material | Fused Silica |
| Coating Material | Inconel 600 |
| Spectral response | Flat |
| Anti-Reflection Coating: (950 nm – 1800 nm) | R<1% |

## 2.3 The IR breadboard

The IR optical bench, built at LESIA for Research & Development purpose in Coronagraphy, was made available for testing the coronagraph prototypes of SPHERE. This facility mimics the optical conditions of SPHERE (F/40 at the focal plane where the masks are installed) and the VLT pupil including the 14% central obscuration and the spiders. The entrance aperture diameter and the Lyot pupil diameter are respectively 18mm and 10mm as in the optical system of SPHERE. The optical components are exclusively reflective to avoid inherent chromatism problems. The artificial on-axis star is obtained from a polychromatic super-continuum laser source. The use of a photonic crystal fiber provides a monomode output from 0.4 to 2.5µm with an average power of 1mW/nm. Several near IR filters with different bandwidth are made available to cover the spectral range between 1.06µm and 2.2µm. The optical design is shown in Figure 4 and Table 3 gives a list of available filters.

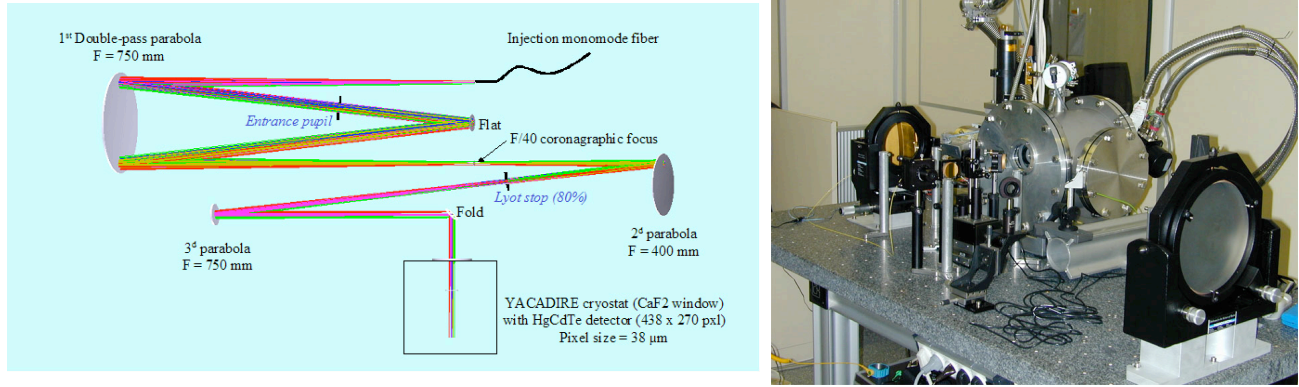

Figure 4: Optical design and picture of the IR breadboard

Table 3: filter list

| Band | Central λ (µm) | HWHM (nm) | Trans. Max (%) |
|---|---|---|---|
| K | 2.2 | 400 (R = 5.3) | 75 |
| H | 1.68 | 240 (R = 7) | 55 |
| H2 | 1.64 | 24 (R = 70) | 48 |
| H3 | 1.59 | 35 (R = 45) | 40 |
| J | 1.191 | 41 (R = 30) | 85 |
| Y | 1.063 | 52 (R = 20) | 88 |

## 3. RESULTS WITH THE HW-4QPM

### 3.1 Contrast performance

Contrast performance of the HW-4QPM has been explored for several conditions. In this paper, we only report on the measurements in the Y, J and H filters. The diffraction by the central obscuration with a 4QPM is a known issue. For that we have compared several Lyot stop designs, some adapted to the diffraction pattern, with a cross-like shape and some circularly symmetric. We found no significant difference in performance indicating that at this level of contrast the central obscuration effect is not dramatic. For simplicity we therefore adopt a circular stop. Figure 5 shows the contrast in the 3 filters assuming this optical setup. In the H band, the contrast reaches $10^{-4}$ at 2.5 λ/D and about $10^{-5}$ at 6λ/D. Performances are slightly degraded at shorter wavelengths since the impact of phase aberrations are more critical. However, the measurements clearly demonstrate the nearly achromatic behavior of the HW-4QPM in the SPHERE wide spectral range. Table 4 compares the result with technical specifications of SPHERE and shows that the HW-4QPM is compliant with the requirements. To quantify the performance we used two metrics, first the peak to peak attenuation on-axis with respect to the non coronagraphic image, and second the PSF wings attenuation which is the local attenuation of the PSF at a given position. Measurements are compared to specifications and to simulations assuming atmospheric seeing corrected with the AO system. In any cases, the performance measured in the IR bench are comparable or better than what can be expected on the sky indicating that intrinsic performance of the coronagraph is suitable to SPHERE. The off-axis transmission (combining mask attenuation and stop throughput) reaches 0.58 at 0.1" and 0.73 at 0.5".

## 3.2 Sensitivity analysis

Sensitivity analysis was made with the HW-4QPM prototype, the VLT-like pupil and the circular stop in the H band. This study was carried out in the mask plane and the Lyot stop plane in X,Y,Z,Θ when relevant. The most stringent parameters are the XY positioning in the 2 planes as reported in Table 5, which need to be maintained with an accuracy better than 50μm. In SPHERE, the XY positioning is achieved with a dedicated Differential Tip Tilt System, which is able to reach an accuracy of 0.5mas (goal 0.2mas). This capability is required by the observing mode which involves a comparison of the data from the meridian while the star is tracked by the system.

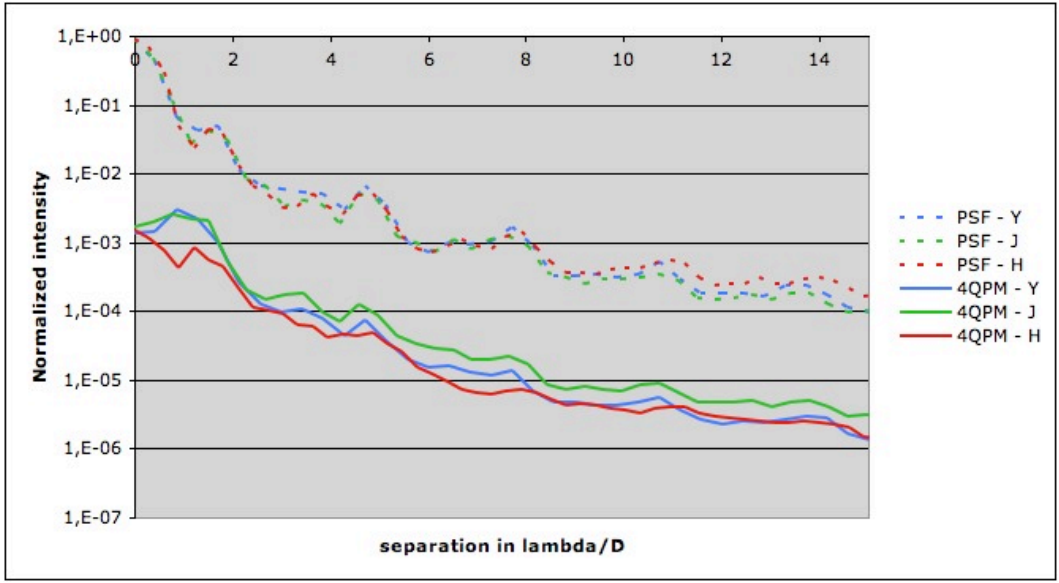

Figure 5: Radial azimuthally averaged contrast in the Y, J and H filters measured with the HW-4QPM prototype.

Table 4: Comparison of measurements with technical specifications for the HW-4QPM

| Band | H (R = 7) | J (R = 20) | Y (R = 30) |
|---|---|---|---|
| Peak to peak attenuation | | | |
| tech specs | >100 | >100 | >50 |
| measured | 600 | 350 | 310 |
| simulation (seeing 0.85'') | 330 | 350 | 300 |
| PSF wings attenuation (0.1-0.5") | | | |
| tech specs | >4 | >4 | >2 |
| measured (0.1'' and 0.5'') | 60 - 85 | 20 - 40 | 60 - 70 |
| simulation (seeing 0.85'') | 120 - 5.0 | 44 - 2.5 | 20 - 1.3 |
| Off axis transmission @ 0.1" | 0.2 (goal: 0.5) | 0.2 (goal:0.5) | 0.2 (goal:0.5) |
| tech specs | | | |
| measured | 0.58 | not measured | not measured |
| Off axis transmission @ 0.5" | 0.4 (goal:0.7) | 0.4 (goal:0.7) | 0.4 (goal:0.7) |
| tech specs | | | |
| measured | 0.73 | not measured | not measured |

Table 5: sensitivity of the HW-4QPM

| parameters | specs |
|---|---|
| *Mask* | |
| IWA | 1.1 λ/D |
| transitions | 1.0 λ/D |
| defocus | < 100μm |
| *Stop* | |
| XY position | < 50μm (each axis) |
| defocus | < 2mm |
| orientation | < 0.5° |

# 4. RESULTS WITH THE APLC

## 4.1 Apodizer transmission profile

The transmission profile of the apodizer is a critical point although the Inconel is known to be grey. We obtained non-coronagraphic pupil images with the apodizer in the beam with the 3 filters and radial azimuthally averaged profiles were plotted against radius (Figure 6). The achromaticity of the apodizer is clearly demonstrated although measurements deviate from the theoretical profile by less than 10% at mid-radius (3 to 7mm). We checked on numerical simulations that performances will not be strongly affected by such a profile as long as we are dealing with $10^4$ to $10^5$ contrast on the coronagraphic image.

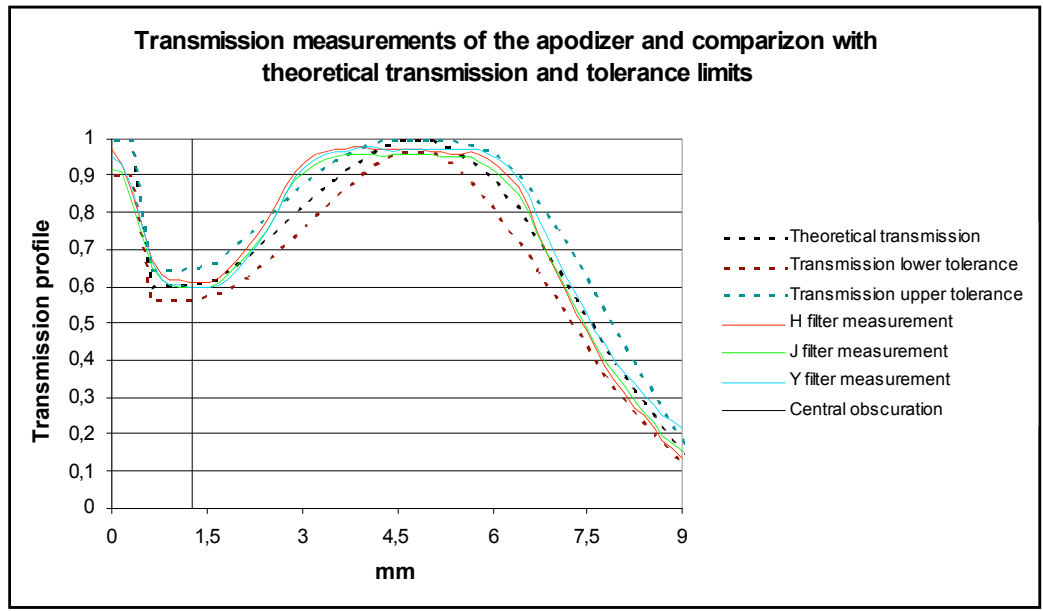

Figure 6: radial transmission profile of the Apodizer prototype.

## 4.2 Contrast performance

As for the HW-4QPM, we measured contrast profiles of the coronagraphic image in the 3 filters with the apodizer prototype and a 4λ/D Lyot mask (at 1.65μm). Again performances are homogeneous although worse in the Y and J bands due to the mismatch between the mask size and the apodizer profile at these wavelengths. However, the chromatism has not a dramatic impact as long as the mask size is set for the longest wavelengths. Performance are reported in Figure 7 and compared to technical specifications in Table 6. As for the HW-4QPM, measurements are better than what we expect on the sky demonstrating the validity of the APLC prototype for SPHERE. The off-axis transmission reaches 0.39 at 0.1" and 0.58 at 0.5" in agreement with specifications.

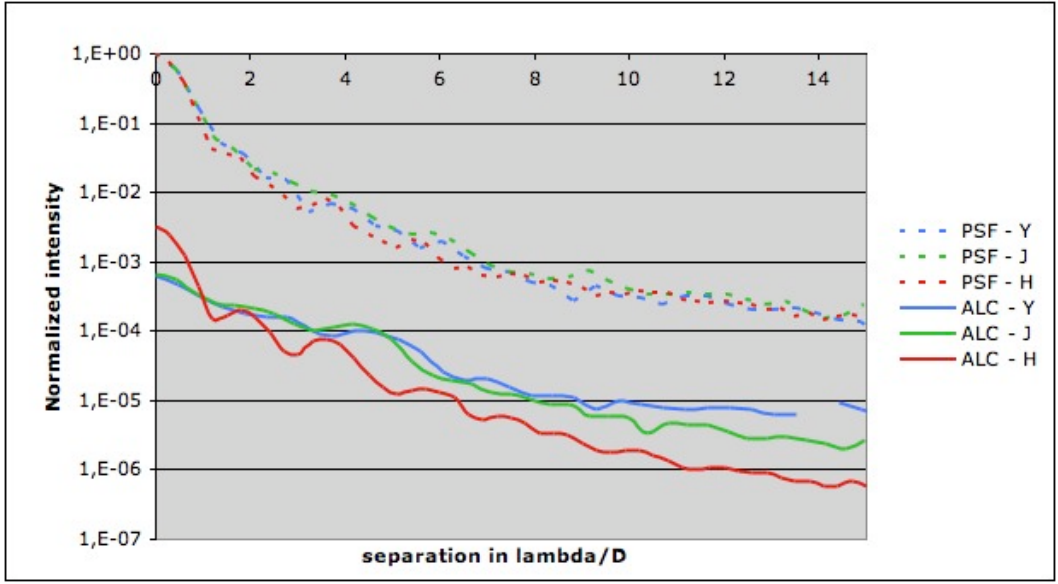

Figure 7: Radial azimuthally averaged contrasts in the Y, J and H filters measured with the APLC prototype.

Table 6: Comparison of measurements with technical specifications for the APLC

| Band | H (R = 7) | J (R = 20) | Y (R = 30) |
|---|---|---|---|
| Peak to peak attenuation | | | |
| tech specs | >100 | >100 | >50 |
| measured | 270 | 830 | 930 |
| simulation (seeing 0.85'') | 160 | 150 | 170 |
| PSF wings attenuation (0.1-0.5") | | | |
| tech specs | >4 | >4 | >2 |
| measured (0.1'' and 0.5'') | 80 - 110 | 30 - 55 | 20 – 10 |
| simulation (seeing 0.85'') | 25 - 1.5 | 10 - 1.1 | 3 - 1.0 |
| Off axis transmission @ 0.1" | 0.2 (goal: 0.5) | 0.2 (goal:0.5) | 0.2 (goal:0.5) |
| tech specs | | | |
| measured | 0.39 | not measured | not measured |
| Off axis transmission @ 0.5" | 0.4 (goal:0.7) | 0.4 (goal:0.7) | 0.4 (goal:0.7) |
| tech specs | | | |
| measured | 0.65 | not measured | not measured |

### 4.3 Sensitivity analysis

The sensitivity analysis of the APLC reveals a lower impact of the mask and stop positioning than for the HW-4QPM a direct consequence of a twice larger Inner Working Angle. This analysis is summarized in Table 7.

Table 7: sensitivity of the APLC

| parameters | specs |
|---|---|
| *Apodizer* | |
| XY position | < 100μm |
| defocus | < 100μm |
| *Mask* | |
| IWA | 2.3 λ/D |
| defocus | < 500μm |
| *Stop* | |
| XY position | < 80μm (each axis) |
| defocus | < 2mm |

**4.4 Improvement of Apodizer transmission profile**

Although, the Apodizer prototype ordered at Reynard Corp. in 2007 has reasonable performance, the transmission profile clearly deviates from the initial specifications. Strong efforts have been made to produce a more compliant prototype. In May 2008, Reynard Corp. has provided us with a new prototype of which the transmission profile is now within specifications. This new component also features a better quality substrate than the former one and it will be presumably tested on the IR coronagraphic bench soon. Figure 8 shows the transmission profile compared to specifications of the new prototype. This final prototyping phase demonstrates the validity of the technique for SPHERE.

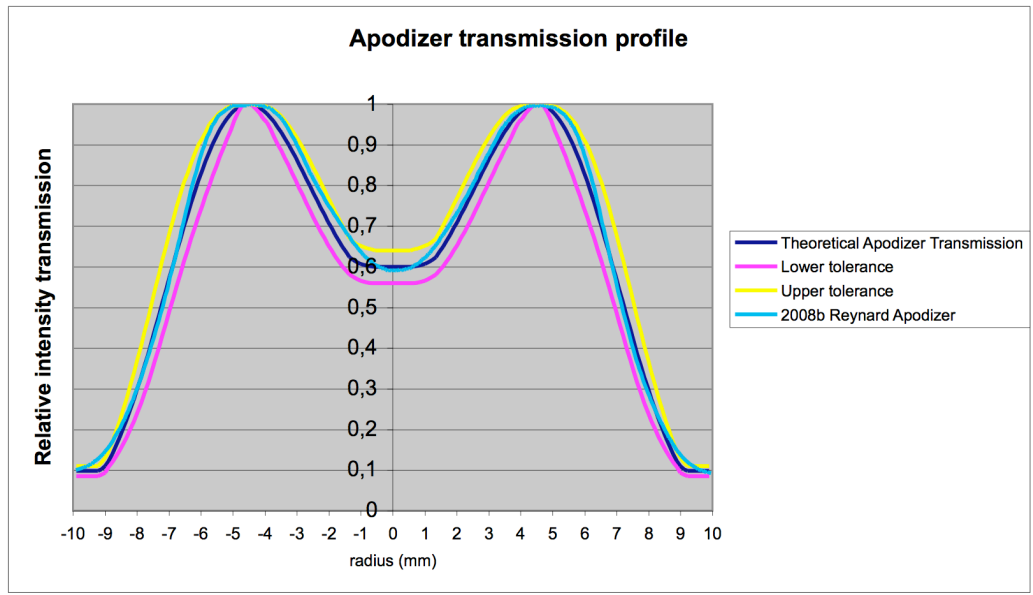

Figure 8: Radial transmission of the 2008 apodizer prototype provided by Reynard Corp.

**5. CONCLUSION**

This paper results of a prototyping phase started almost 2-years ago in the context of the SPHERE phase B. This work carried out by FIZEAU and LESIA was very fruitful and has clearly demonstrated the feasibility of manufacturing two achromatic coronagraphs : the Half-Waveplates 4QPM and the Apodized Pupil Lyot Coronagraph. These prototypes are now within the specifications derived from numerical simulations and achieve the requirements in terms of contrast and achromaticity. In addition, they are representative of the final components to be installed in SPHERE (size, thickness,

A/R coatings, …) and were also tested at a scale identical to that of SPHERE (F/ratio, pupil/stop diameters). It reinforces our confidence in the performance of these two coronagraphs. The IR bench facility has allowed us to measure directly the sensitivities of the coronagraphic elements in terms of positioning that we compared with our previous estimates based on simulation. This particular work was extremely important to drive some requirements of the whole system, like the presence of the Pupil tip/tilt mirror and that of a Differential Tip Tilt Sensor.